\documentclass[aps,pra,twocolumn,superscriptaddress]{revtex4-2}
\usepackage{mathtools}
\usepackage{graphicx}
\usepackage{hyperref}
\usepackage{enumerate}

\begin{document}
\title{Distinguishing thermal and pseudothermal light by testing the Siegert relation}

\author{Xi~Jie~Yeo}
\affiliation{Centre for Quantum Technologies, National University of Singapore, 3 Science Drive 2, Singapore 117543}

\author{Justin~Yu~Xiang~Peh}
\affiliation{Centre for Quantum Technologies, National University of Singapore, 3 Science Drive 2, Singapore 117543}

\author{Darren~Ming~Zhi~Koh}
\affiliation{Centre for Quantum Technologies, National University of Singapore, 3 Science Drive 2, Singapore 117543}

\author{Christian~Kurtsiefer}
\affiliation{Centre for Quantum Technologies, National University of Singapore, 3 Science Drive 2, Singapore 117543}
\affiliation{Department of Physics, National University of Singapore, 2 Science Drive 3, Singapore 117551}
\email[]{christian.kurtsiefer@gmail.com}

\author{Peng~Kian~Tan}
\affiliation{Centre for Quantum Technologies, National University of Singapore, 3 Science Drive 2, Singapore 117543}

\date{\today}

\begin{abstract}
Thermal light, including blackbody radiation and spontaneous emission,
exhibits photon bunching.
Thermal light sources, however, typically yield low spectral densities,
limiting their practical utility. Pseudothermal light sources with higher
brightness and longer coherence time are often employed instead.
While pseudothermal light also exhibits photon bunching, this property
may not suffice to fully replicate the behavior of genuine thermal light. Here we demonstrate a method to directly test the Siegert relation for two sources of photon-bunched light, laser light scattered from a rotating ground glass and spontaneously emitted light from a gas discharge lamp, probing a fundamental criterion expected of thermal light.
\end{abstract}

\maketitle

\section{Thermal and pseudothermal light}

Blackbody radiation forms stationary light fields across a broad
frequency spectrum without fixed phase relationships between light at different
locations or times~\cite{Davis1996}.
Similarly, light with random phases between individual emission processes is produced when spontaneous emission arises from an ensemble of particles, like atoms, with motion in thermal equilibrium~\cite{Planck1900,Einstein1905,Glauber1963_PhotonCorrelations,Loudon2000}.

Thermal light, including blackbody radiation~\cite{PK2016,Deng2019} and
spontaneous emission from gas discharge
lamps~\cite{Rebka1957,HBT1958,Morgan1966,Scarl1968}, exhibits photon
bunching~\cite{HBT1954,HBT1956,Glauber1963_QuantumTheory,MandelWolf1995,Fox2006},
where the photodetection events appear closer together than as described by
random Poissonian timing statistics~\cite{Glauber1963}; mathematically, this
is expressed by a second order correlation function $g^{(2)}(\tau=0)>1$.

However, the spectral density of blackbody radiation is determined by its surface temperature, which limits its photon bunching output brightness below practical thresholds. The broadband nature of the blackbody spectrum shortens the coherence timescale of the bunching behavior to the femtoseconds regime, rendering its photon bunching inaccessible to standard photodetectors.

Alternative “pseudothermal” light sources are therefore commonly employed to demonstrate photon bunching. 
Typical implementations include laser beams scattered from time‑varying dispersive media, such as rotating ground glass diffusers~\cite{MS1964,Arecchi1965,Gonsiorowski1983,Scarcelli2004,Valencia2005,Zhou2017}, or particles undergoing Brownian motion suspended in a liquid~\cite{Arecchi1967,Pecora1985,Dravins2015,PK2017}.
Such pseudothermal light sources generate bright photon-bunched light owing to the underlying laser excitation, with coherence timescales long enough to be readily resolved by available photodetectors.

Here, we demonstrate that pseudothermal light can yield photon correlations distinguishable from those of thermal radiation, even though both exhibit the temporal photon bunching characteristic.

\section{Signature photon bunching}
\begin{figure}
  \begin{center}
    \includegraphics[width=\columnwidth]{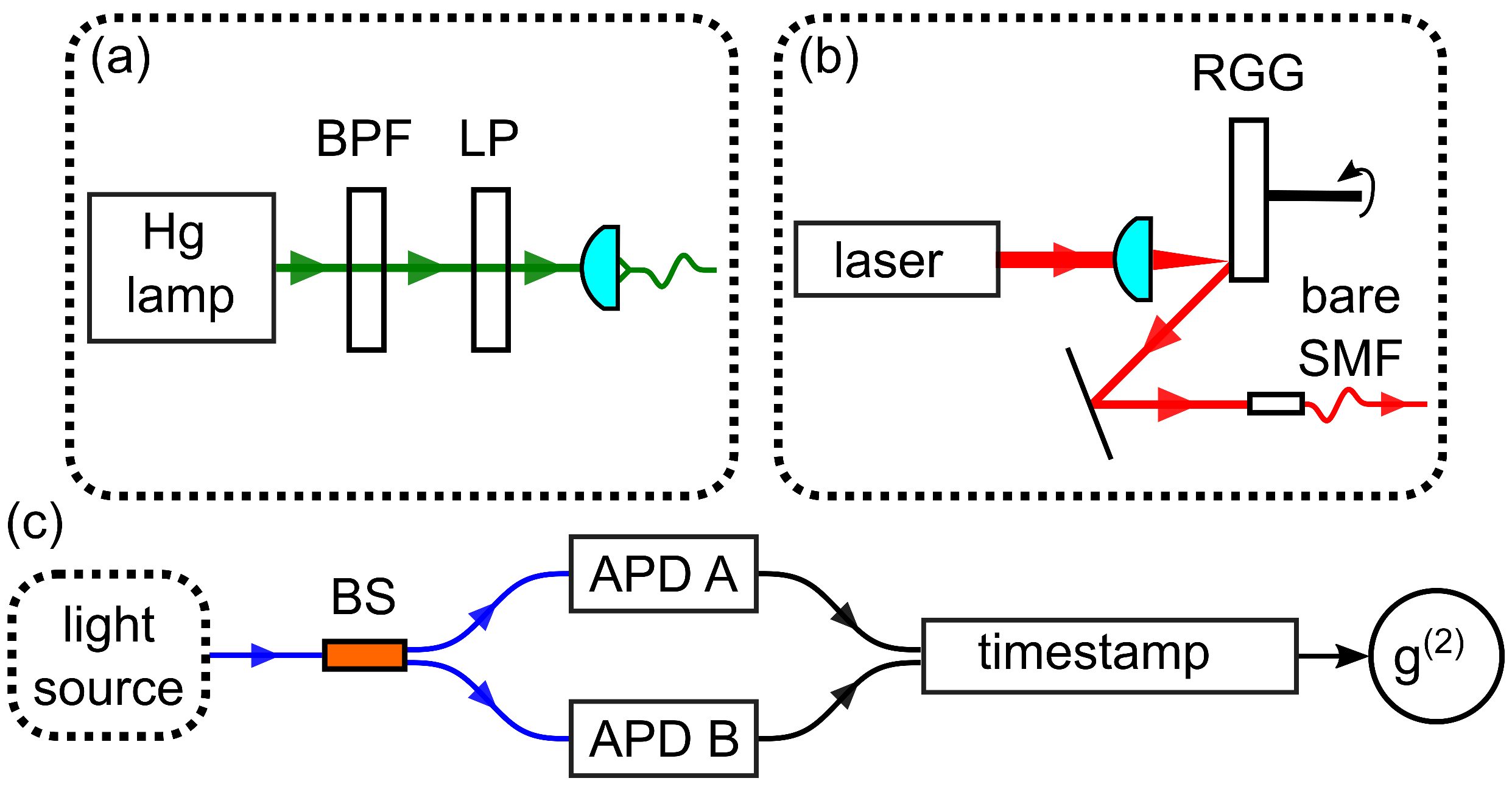}
  \end{center}
  \caption{
    (a) Thermal light source based on a Mercury (Hg) vapor gas discharge lamp,
    (b) pseudothermal light generated by laser light scattered off a rotating
    ground glass diffuser. 
    (c) Hanbury-Brown--Twiss interferometer to measure the second-order
    photodetection time correlations $g^{(2)}(\tau)$. 
    (BPF: bandpass filter, LP: linear polariser, RGG: rotating ground glass,
    BS: fibre-based beamsplitter, APD: avalanche photodetectors, SMF: single
    mode fibre)
  } \label{fig:g2_setup}
\end{figure}

We compare two photon-bunched light sources: a thermal light source based on
spontaneous emissions from a Mercury (Hg) low pressure gas discharge lamp, and a
pseudothermal light source based on laser light scattered off a rotating
ground glass plate (see Figs.\,\ref{fig:g2_setup}(a) and (b)).

For thermal light, the spontaneous emission around 546.1\,nm from the Hg lamp is selected by a 546\,nm spectral filter with a 3\,nm optical passband.
Multiple spectral emission lines due to hyperfine splitting resulting from a
natural mix of Hg isotopes with non-zero nuclear spins are captured.
The presence of multiple emissions that are mutually incoherent reduce the
resulting $g^{(2)}(0)$ from the ideal thermal light value of 2.
The spectrally filtered light is transmitted through a linear polariser to increase interferometric visibility, and projected into a single spatial mode by a single mode optical fibre.

Pseudothermal light is prepared by scattering light from a 780\,nm distributed
feedback laser focused on a rotating ground glass difffuser of grit 1500
(see Fig.\,\ref{fig:g2_setup}(b). 
The bare tip of an optical fiber (single mode for 780\,nm) is positioned about 19\,cm away from the illuminated spot to collect light scattered from the ground glass.
Light scattered from a rotating ground glass is expected to show photon
bunching described by a second order correlation function $g^{(2)}(\tau)$ with a
Gaussian profile~\cite{Estes1971,Jakeman1975,Pusey1976,Kuusela2017},
\begin{equation}
  g^{(2)}(\tau) = 1 +
  e^{-\left(\frac{\tau}{\tau_{\text{RGG}}}\right)^{2}}\,,
\end{equation}
where $\tau_{\text{RGG}}$ is the coherence timescale of the photon bunching
signature of this souce.

The focal spot size $\phi$ of the beam on the ground glass surface is about
4\,$\mu$m in diameter, and positioned at a radial distance $R$ of about 10\,mm
from the rotation axis. A motor rotates the ground glass with a period $T\approx4$\,ms.
With the scattered light collected at a distance of 19\,cm away, which is
significantly larger than the spot size of $\phi=4\,\mu$m, the  coherence
timescale $\tau_{\text{RGG}}$ is
predicted~\cite{Estes1971,Jakeman1975,Pusey1976,Kuusela2017} as 
$\tau_{\text{RGG}} \approx \phi T/(4 \pi R)$, or
$\tau_{\text{RGG}}\approx130$\,ns for our parameters.

\begin{figure}
  \begin{center}
    \includegraphics[width=\columnwidth]{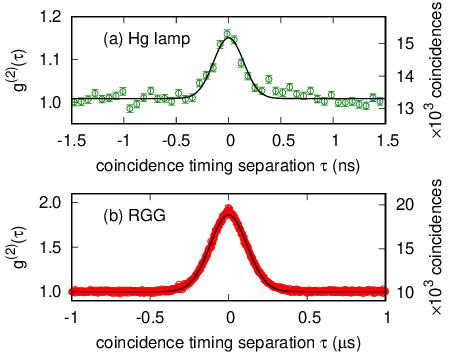}
  \end{center}
 \caption{Photon bunching signatures. (a) The Hg lamp exhibits photon bunching
   with $g_{\text{Hg}}^{(2)}(\tau=0)=1.142\pm0.007$ and coherence timescale
   $\tau_{\text{Hg}}=0.35\pm0.02\,$ns (both numeric values are results of fits, solid
   line). 
   (b) Results for scattered light from the rotating ground glass yield
   $g_{\text{RGG}}^{(2)}(0)=1.859\pm0.002$ and $\tau_{\text{RGG}}=277.5\pm0.6\,$ns. 
} \label{fig:g2}
\end{figure}

To detect for photon bunching signatures in light from the two sources, a Hanbury-Brown--Twiss (HBT) interferometer (see Fig.\,\ref{fig:g2_setup}(c) is
used. For this, light from a source is sent through a beamsplitter to illuminate a pair of actively quenched Silicon avalanche photodiodes (APD).
The photo-detection events were then timestamped, from which detection time
differences between the detectors were histogrammed, and histograms were fitted with Gaussian models as illustrated in Fig.\,\ref{fig:g2}.

Light from both the Hg lamp and the rotating ground glass exhibit temporal photon
bunching. The fitted peak $g_{\text{Hg}}^{(2)}(\tau=0)=1.142\pm0.007$ is significantly lower than the
theoretical thermal light peak $g^{(2)}(\tau=0)=2$, as anticipated from the
contribution of different spectral lines. Furthermore, the short coherence
timescale of $\tau_{\text{Hg}}=0.35\pm0.02\,$ns makes it difficult for standard
photodetectors to resolve the timing of photo-detection pairs.

In contrast, the output from the rotating ground glass exhibits a noticeably
stronger photon bunching behaviour with $g_{\text{RGG}}^{(2)}(\tau=0)=1.859\pm0.002$, and a
coherence timescale $\tau_{\text{RGG}}=277.5\pm0.6\,$ns that is 3 orders longer, and
thus straightforward to resolve with readily available detectors. This makes
scattering light from a rotating ground glass a popular pseudothermal light
source to demonstrate photon bunching.

\section{Testing the Siegert relation}
Thermal light exhibits photon timing correlations, including the characteristic photon bunching, that satisfy the Siegert relation~\cite{Siegert1943,Pike1974,Jakeman1980,Lemieux1999,Ferreira2020,Las2022}
\begin{equation}
  g^{(2)}(\tau) = 1 + \lvert g^{(1)}(\tau) \rvert ^{2}\,,
  \label{eqn:siegert}
\end{equation}
where $g^{(1)}(\tau)$ is the first order correlation function, and its modulus $\lvert g^{(1)}(\tau) \rvert$ is the interferometric visibility.

The $g^{(2)}(\tau)$ can be determined by photon bunching measurements in a HBT interferometer as shown in the previous section, and is adopted as a general test for thermal light due to its practical ease and convenience.

The $g^{(1)}(\tau)$ can be obtained with a Michelson interfereometer scanning
over a range of optical path length differences. This can be difficult as the
scanning range needs to be on the order of the coherence timescale. For the
example above with $\tau_{\text{RGG}}=277.5\,$ns, this would necessitate an
impractical scan over 83\,metres of path length difference.

\begin{figure}
  \begin{center}
    \includegraphics[width=\columnwidth]{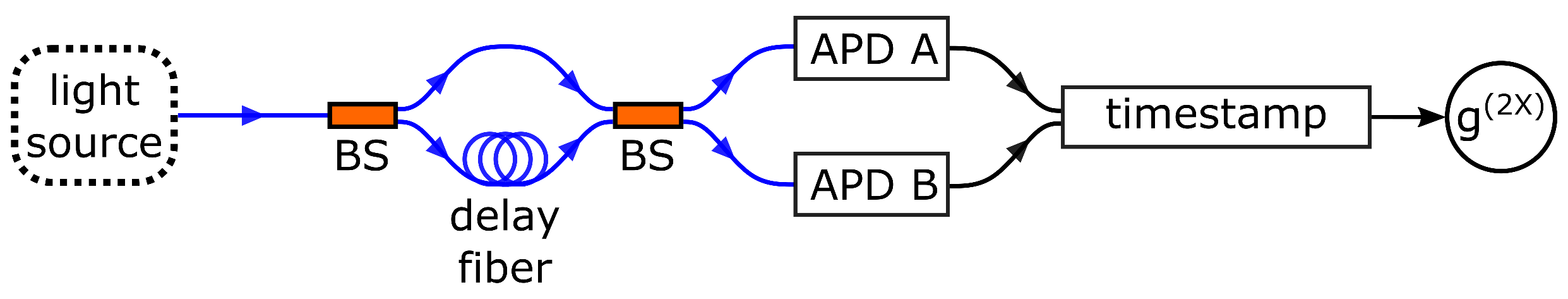}
  \end{center}
  \caption{Experimental setup for measuring interferometric photon correlations $g^{(2X)}(\tau)$. 
The delay fibers are chosen to introduce an optical delay of
$\Delta=2.22\,\mu$s for testing scattered light from the rotating ground
glass, and $\Delta=10\,$ns for testing light from the Hg lamp.
(BS: Beamsplitter, APD: avalanche photodetectors)
    } \label{fig:g2X_setup}
\end{figure}

Alternatively, an asymmetric Mach-Zehnder interferometer can be used to test
the Siegert relation via an interferometric correlation $g^{(2X)}(\tau)$ (see Fig.\,\ref{fig:g2X_setup}). This reveals both the second order correlation function $g^{(2)}(\tau)$ and the modulus of the first order correlation function $\lvert g^{(1)}(\tau) \rvert$ simultaneously.
This method has the advantage of being phase independent by measuring $\lvert
g^{(1)}(\tau) \rvert$ instead of $g^{(1)}(\tau)$, and does not require
a range of optical path length differences.

The respective light fields $E_{A,B}(t)$ at the output ports $A,B$ of an intensity-balanced interferometer are 
\begin{equation}
    E_{A,B}(t) = \frac{E(t) \pm E(t+\Delta)}{\sqrt{2}}\,,
    \label{eqn:fields}
\end{equation}
where $E(t)$ is the input light field of the interferometer.
The fixed optical delay $\Delta$ is introduced by a single-mode optical fiber,
chosen to be much larger than the expected characteristic coherence time of
the light.

Photoevents in our implementation are detected by Silicon avalanche photodetectors (APDs) at the output ports of the interferometer and timestamped.
The interferometric photon correlation $g^{(2X)}(\tau)$ between the
photodetectors is
\begin{equation} 
    \begin{aligned}
        g^{(2X)}(\tau) = &\frac{\langle E^{*}_{A}(t+\tau)E^{*}_{B}(t)E_{B}(t)E_{A}(t+\tau)\rangle}{ \langle E^{*}_{A}(t)E_{A}(t)\rangle \langle E^{*}_{B}(t)E_{B}(t)\rangle}\\
    \end{aligned}
  \label{eqn:g2X}
\end{equation}
where $\langle\dots\rangle$ is the ensemble average over measurement time $t$, and $\tau$ is the two-photoevent coincidence timing separation.

Upon expansion of Eq.\,(\ref{eqn:g2X}) using Eq.\,(\ref{eqn:fields}), it can be shown~\cite{lebreton2013exp,XJ2023} that the non-zero terms are
\begin{equation} 
    \begin{aligned}
g^{(2X)}(\tau) = &\frac{1}{4}\,g^{(2)}(\tau+\Delta)+ \frac{1}{4}\,g^{(2)}(\tau-\Delta) \\
+ &\frac{1}{2}[g^{(2)}(\tau) -  |g^{(1)}(\tau)|^{2}]\,.
    \end{aligned}
  \label{eqn:zerotime}
\end{equation}
The temporal photon bunching $g^{(2)}(\tau=0)$ signature would now manifest into two side
peaks $g^{(2X)}(\tau)$ at $1/4$ of its original amplitude, positioned at
photo-detection time delays $\tau=\pm\Delta$.

If the test light obeys the Siegert relation Eq.\,(\ref{eqn:siegert}),
and recalling that the optical delay length is chosen
such that $g^{(2)}(\tau=\Delta)=1$, then
$g^{(2X)}(\tau=0)=\frac{1}{4}+\frac{1}{4}+\frac{1}{2}=1$. This results in a flat
coincidence floor around $\tau=0$.

\begin{figure}
  \begin{center}
    \includegraphics[width=\columnwidth]{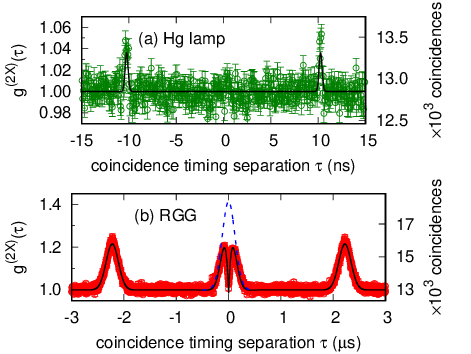}
  \end{center}
 \caption{Experimental results for interferometric photon correlations. (a)
   Light from the Hg lamp shows two bunching side peaks
   $g_{\text{Hg}}^{(2X)}(\tau=\pm\Delta)=1.044\pm0.003$ at reduced amplitudes, with
   $\tau_{\text{Hg}}=0.41\pm0.03\,$ns. 
   (b) Scattered light from the rotating ground glass exhibits a central bunching peak (blue dashed lines) $g_{\text{RGG}}^{(2X)}(\tau=0)=1.412\pm0.002$ and two smaller side peaks $g_{\text{RGG}}^{(2X)}(\tau=\pm\Delta)=1.206\pm0.002$, with $\tau_{\text{RGG}}=172.8\pm0.6\,$ns. 
   The solid line shows a fit to Eq.\,(\ref{eqn:zerotime}).
} \label{fig:g2X}
\end{figure}

Experimental results of $g^{(2X)}(\tau)$ for the two light sources
tested in this work are shown in Fig.\,\ref{fig:g2X}. The flat region near
$\tau=0$ for light emitted by the Hg lamp suggests good agreement with the Siegert
relation, and therefore compatible with expectations for thermal light.

Light prepared by the rotating ground glass setup shows
a non-flat bunching feature around $\tau=0$, clearly violating the Siegert
relation ($g^{(2)}(\tau) -|g^{(1)}(\tau)|^{2}\neq1$, see Fig.\,\ref{fig:g2X} (b)).
The fit of the experimental $g_{\text{RGG}}^{(2X)}(\tau)$ to Eq.\,(\ref{eqn:zerotime})
yields a central bunching peak $g_{\text{RGG}}^{(2X)}(0)=1.412\pm0.002$ and
$-|g^{(1)}(0)|^{2}=-0.421\pm0.003$. However, although the amplitude of
$|g^{(1)}(0)|^{2}$ is similar to the central bunching peak amplitude, its
coherence timescale $\tau_{c}=149\pm2\,$ns is shorter than $\tau_{\text{RGG}}=172.8\pm0.6\,$ns, and
so it does not fully cancel out the bunching feature around $\tau=0$.

This deviation from the
expectation for thermal light suggests the rotating ground glass source to be
considered a pseudothermal light source, and not a thermal light source.

\section{Conclusion}
We find that the test of the Siegert relation can be a useful tool to
distinguish thermal from pseudothermal light, providing additional information
beyond the observation of photon bunching. Therefore, probing for the
interferometric photon correlation $g^{(2X)}(\tau)$ can be a stronger witness
of thermal light.

\begin{acknowledgments}
This research is supported by the Quantum Engineering Programme through NRF2021-QEP2-03-P02, the Ministry of Education, and the National Research Foundation, Prime Minister's Office, Singapore.
\end{acknowledgments}

\bibliography{references}
\end{document}